\documentclass[aps,twocolumn,amsmath,amssymb,superscriptaddress,prb]{revtex4}

\usepackage{graphicx}
\usepackage{bm}
\usepackage[dvips]{color} 

\begin{document}

\title{Spin-current emission governed by nonlinear spin dynamics}

\author{Takaharu Tashiro}
\affiliation{Department of Applied Physics and Physico-Informatics, Keio University, Yokohama 223-8522, Japan}

\author{Saki Matsuura} \affiliation{Department of Applied Physics and Physico-Informatics, Keio University, Yokohama 223-8522, Japan}

\author{Akiyo Nomura} \affiliation{Department of Applied Physics and Physico-Informatics, Keio University, Yokohama 223-8522, Japan}

\author{Shun Watanabe}
\affiliation{Cavendish Laboratory, University of Cambridge, J. J. Thomson Avenue, Cambridge CB3 0HE, United Kingdom}

\author{Keehoon Kang}
\affiliation{Cavendish Laboratory, University of Cambridge, J. J. Thomson Avenue, Cambridge CB3 0HE, United Kingdom}

\author{Henning Sirringhaus}
\affiliation{Cavendish Laboratory, University of Cambridge, J. J. Thomson Avenue, Cambridge CB3 0HE, United Kingdom}

\author{Kazuya Ando\footnote{Correspondence and requests for materials should be addressed to ando@appi.keio.ac.jp}}
\affiliation{Department of Applied Physics and Physico-Informatics, Keio University, Yokohama 223-8522, Japan}
\affiliation{PRESTO, Japan Science and Technology Agency, Kawaguchi, Saitama 332-0012, Japan}

\begin{abstract}
Coupling between conduction electrons and localized magnetization is responsible for a variety of phenomena in spintronic devices. This coupling enables to generate spin currents from dynamical magnetization. Due to the nonlinearity of magnetization dynamics, the spin-current emission through the dynamical spin-exchange coupling offers a route for nonlinear generation of spin currents. Here, we demonstrate spin-current emission governed by nonlinear magnetization dynamics in a metal/magnetic insulator bilayer. The spin-current emission from the magnetic insulator is probed by the inverse spin Hall effect, which demonstrates nontrivial temperature and excitation power dependences of the voltage generation. The experimental results reveal that nonlinear magnetization dynamics and enhanced spin-current emission due to magnon scatterings are triggered by decreasing temperature. This result illustrates the crucial role of the nonlinear magnon interactions in the spin-current emission driven by dynamical magnetization, or nonequilibrium magnons, from magnetic insulators. 
\end{abstract}

\maketitle

Dynamical magnetization in a ferromagnet emits a spin current,~\cite{Tserkovnyak1,Brataas} enabling to explore the physics of spin transport in metals and semiconductors.~\cite{PhysRevLett.111.106601,PhysRevLett.107.066604,PhysRevB.87.174417,PhysRevLett.111.176601,PhysRevB.88.014404,Fertspinpump,PhysRevLett.106.216601,Hujun,Saitoh,PhysRevLett.110.127201,AndoNMad,KurebayashiNM,PhysRevB.86.134419,Chumakmt,benjamintemp,AndoShun,jungfleisch:022411,PhysRevApplied.1.044004,Andobistable,PhysRevLett.111.247202} The dynamical spin-current emission has been achieved utilizing ferromagnetic metals, semiconductors, and insulators.~\cite{Mizukami,PhysRevLett.107.046601,GaMnSP,kajiwara} In particular, the discovery of the spin-current emission from a magnetic insulator yttrium iron garnet, Y$_3$Fe$_5$O$_{12}$, has drawn intense experimental and theoretical interests, opening new possibilities to spintronics based on metal/insulator hybrids, where angular momentum can be carried by both electrons and magnons.

\begin{figure*}[tb]
\includegraphics[scale=1]{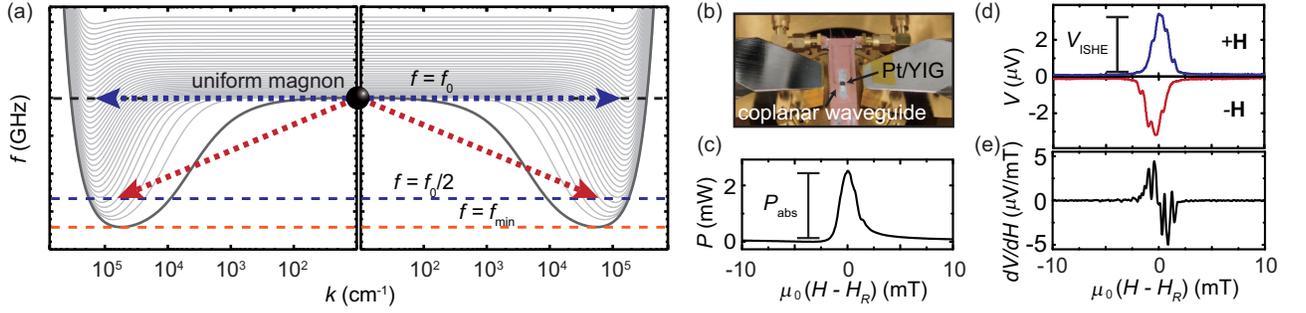}
\caption{{\bfseries Detection of spin-current emission.} (a) The magnon dispersion in Y$_3$Fe$_5$O$_{12}$, where $f$ and $k$ are the frequency and wavenumber of magnons, respectively. The dispersion of the first 40 thickness modes propagating along and opposite to the magnetic field is shown. The blue and red arrows represent the four and three magnon scatterings. The magnon dispersion shows that both the three and four magnon scatterings create secondary magnons with small group velocity. The lowest frequency is $f=f_\text{min}$. (b) The experimental setup. The Pt/Y$_3$Fe$_5$O$_{12}$ film placed on the coplanar waveguide was cooled using a Gifford-McMahon cooler. (c) Magnetic field ($H$) dependence of the microwave absorption $P$ for the Pt/Y$_3$Fe$_5$O$_{12}$ film at $f_0= 7.6$ GHz and $P_\text{in}=10$ mW. $\mu_0 H_R=183$ mT is the resonance field. $P_\text{abs}$ is the definition of the magnitude of the microwave absorption intensity. The absorption peak structure comprises multiple signals due to spin-wave modes. (d) $H$ dependence of the electric voltage $V$. $V_\text{ISHE}$ is the magnitude of the electric voltage. The blue and red data were measured with the in-plane magnetic field ${\bf H}$ and $-{\bf H}$, respectively. (e) $H$ dependence of  $dV(H)/dH$ for the Pt/Y$_3$Fe$_5$O$_{12}$ film. The damping constant of the Pt/Y$_3$Fe$_5$O$_{12}$ film was roughly estimated to be $5\times 10^{-4}$ from $f_0$ dependence of the linewidth at 5 mW.}
\label{fig1} 
\end{figure*}

A ferrimagnetic insulator yttrium iron garnet, Y$_3$Fe$_5$O$_{12}$, is characterized by the exceptionally small magnetic damping, making it a key material for the development of the physics of nonlinear magnetization dynamics.~\cite{BLS,Serga,Lenk2011107} The nonlinear magnetization dynamics in Y$_3$Fe$_5$O$_{12}$ has been extensively studied both experimentally and theoretically in the past half a century, benefited by the exceptional purity, high Curie temperature, and simplicity of the low-energy magnon spectrum.~\cite{Serga,Lenk2011107,patton,Turbulence} Recently, the nonlinear magnetization dynamics has been found to affect the spin-current emission from the magnetic insulator; the spin-current emission is enhanced by magnon scattering processes [see Fig.~\ref{fig1}(a)], triggered by changing the excitation frequency or power of the magnetization dynamics.~\cite{KurebayashiNM,PhysRevB.86.134419,SakimuraNC} These findings shed new light on the long-standing research on nonlinear magnetization dynamics, promising further development of spintronics and magnetics based on the magnetic insulator.

In this work, we demonstrate that the spin-current emission from Y$_3$Fe$_5$O$_{12}$ is strongly affected by nonlinear magnetization dynamics at low temperatures. The spin-current emission is probed by the inverse spin Hall effect (ISHE) in a Pt film attached to the Y$_3$Fe$_5$O$_{12}$ film,~\cite{Saitoh,KimuraPRL,Valenzuela} which enables to measure temperature dependence of the spin-current emission from the magnetic insulator under various conditions. In spite of the simple structure of the metal/insulator bilayer, we found nontrivial variation of the spin-current emission; the temperature dependence of the spin-current emission strongly depends on the microwave frequency and excitation power. This result reveals that nonlinear spin-current emission due to three and four magnon scatterings emerges by decreasing temperature, even at constant magnon excitation frequency and power. This finding provides a crucial piece of information for understanding the spin-current emission from ferromagnetic materials and investigating the magnon interactions in the metal/insulator hybrid.

A single-crystal Y$_3$Fe$_5$O$_{12}$ (111) film ($3\times 5$ mm$^{2}$) with a thickness of 5 $\mu$m was grown on a Gd$_{3}$Ga$_{5}$O$_{12}$ (111) substrate by liquid phase epitaxy (purchased from Innovent e.V., Jena). After the substrates were cleaned by sonication in deionized water, acetone and isopropanol, a piranha etching process, a mixture of H$_{2}$SO$_{4}$ and H$_{2}$O$_{2}$ (with the ratio of 7 : 3), was applied, then to be able to remove any residuals an oxygen plasma cleaning was performed outside a sputtering chamber. On the top of the film, a 10-nm-thick Pt layer was sputtered in an Ar atmosphere. Prior to sputtering 10-nm-thick Pt layer, an argon plasma cleaning was also performed {\it in-situ}. The Pt/Y$_3$Fe$_5$O$_{12}$ bilayer film was placed on a coplanar waveguide, where a microwave was applied to the input of the signal line as show in Fig.~\ref{fig1}(b). Two electrodes were attached to the edges of the Pt layer. The signal line is 500 $\mu$m wide and the gaps between the signal line and the ground lines are designed to match to the characteristic impedance of 50 $\Omega$. An in-plane external magnetic field ${\bf H}$ was applied parallel to the signal line, or perpendicular to the direction across the electrodes.~\cite{Saitoh} Figure~\ref{fig1}(c) shows the in-plane magnetic field $H$ dependence of the microwave absorption $P$ measured by applying a 10 mW microwave with the frequency of $f_0=7.6$ GHz at $T=300$ K. Under the ferromagnetic resonance condition $H=H_R$, dynamical magnetization in the Y$_3$Fe$_5$O$_{12}$ layer emit a spin current $j_s$ into the Pt layer, resulting in the voltage generation through the ISHE as shown in Fig.~\ref{fig1}(d).~\cite{Tserkovnyak1,Brataas} The sign of the voltage is changed by reversing ${\bf H}$, consistent with the prediction of the spin-current emission from the magnetic insulator.~\cite{AndoJAPfull} Here, the absorption spectrum comprises multiple resonance signals due to spin-wave modes, including magnetostatic surface waves and backward-volume magnetostatic waves in addition to the ferromagnetic resonance. To extract the damping constant for the Pt/Y$_3$Fe$_5$O$_{12}$ film, we have plotted $dV/dH$ in Fig.~\ref{fig1}(e), which allows rough estimation of the damping constant, $\alpha\sim5\times 10^{-4}$.

\begin{figure}[tb]
\includegraphics[scale=1]{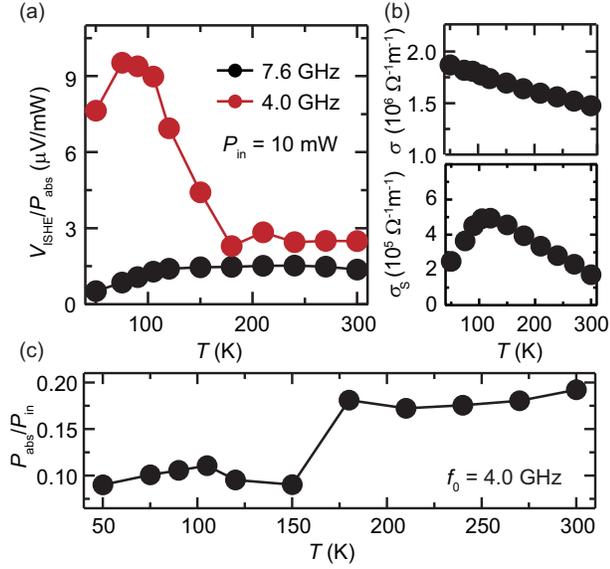}
\caption{{\bfseries Temperature evolution of spin-current emission.} (a) Temperature ($T$) dependence of $V_\text{ISHE}/P_\text{abs}$ for the Pt/Y$_3$Fe$_5$O$_{12}$ film at $f_0=7.6$ (the black circles) and 4.0 GHz (the red circles). The data were measured with $P_\text{in}=10$ mW microwave excitation. (b) $T$ dependence of the electrical conductivity $\sigma$ and the spin Hall conductivity $\sigma_\text{s}$ for the Pt/Y$_3$Fe$_5$O$_{12}$ film. (c) $T$ dependence of $P_\text{abs}/P_\text{in}$, where $P_\text{abs}$ is the microwave absorption intensity, for $P_\text{in}=10$ mW and $f_0=4.0$ GHz.}
\label{fig2} 
\end{figure}

Figure~\ref{fig2}(a) shows temperature dependence of $V_\text{ISHE}/P_\text{abs}$, where $V_\text{ISHE}$ and $P_\text{abs}$ are the magnitude of the microwave absorption and electric voltage, respectively; $V_\text{ISHE}/P_\text{abs}$ characterizes the angular-momentum conversion efficiency from the microwaves into spin currents. Notably, $V_\text{ISHE}/P_\text{abs}$ increases drastically below $T=150$ K by decreasing $T$ at $f_0=4.0$ GHz. This drastic change is irrelevant to the temperature dependence of the spin pumping and spin-charge conversion efficiency in the Pt/Y$_3$Fe$_5$O$_{12}$ bilayer, such as the spin Hall angle $\theta_\text{SHE}$, the spin pumping conductance $g_\text{eff}$, the spin diffusion length $\lambda$, and the electrical conductivity $\sigma$. Figure~\ref{fig2}(b) shows the temperature dependence of the electrical conductivity $\sigma$ and the spin Hall conductivity $\sigma_\text{s}$. The spin Hall conductivity was obtained from the temperature dependence of $V_\text{ISHE}/P_\text{abs}$ at 10 mW for $f_0=7.6$ GHz shown in Fig.~\ref{fig2}(a); the value of $V_\text{ISHE}/P_\text{abs}$ is insensitive to the excitation power from 5 to 15 mW, indicating that the spin-current emission is reproduced with a liner spin-pumping model:~\cite{iguchiJJAP}
\begin{equation}
\frac{{{V_{{\rm{ISHE}}}}}}{{{P_{{\rm{abs}}}}}} = \frac{{2e{w_{\rm{F}}}{\sigma _{\rm{s}}}f_0\lambda {g_{{\rm{eff}}}}\tanh (d/2\lambda )}}{{{\mu _0}{\sigma ^2}d{v_{\rm{F}}}{M_{\rm{s}}}\Delta H\sqrt {{{(\gamma {\mu _0}{M_\text{s}})}^2} + {{(4\pi f_0)}^2}} }},
\end{equation}
where $w_\text{F}=3.0$ mm and $v_\text{F}=7.5\times10^{-11}$ m$^3$ are the width and volume of the Y$_3$Fe$_5$O$_{12}$ film. $d=10$ nm is the thickness of the Pt layer. $\mu_0\Delta H$ is the half-maximum full-width of the ferromagnetic resonance linewidth. For the calculation of $\sigma_\text{s}$, we used the measured parameters of the electrical conductivity $\sigma$ and saturation magnetization $M_\text{s}$. The spin-diffusion length~\cite{nakayama2012geometry} $\lambda=7.7$ nm and spin pumping conductance~\cite{0295-5075-96-1-17005} $g_\text{eff}=4.0\times 10^{18}$ m$^{-2}$ were assumed to be independent of temperature, as demonstrated previously.~\cite{4885086} The spin Hall conductivity of the Pt layer shown in Fig.~\ref{fig2}(b) increases with decreasing temperature above 100 K. Below 100 K, the spin Hall conductivity decreases with decreasing temperature. This feature is qualitatively consistent with the previous report.~\cite{4885086} Although the spin Hall conductivity varies with temperature, the variation of the spin Hall conductivity alone is not sufficient to explain the drastic increase of $V_\text{ISHE}/P_\text{abs}$ for $f_0=4$ GHz shown in Fig.~\ref{fig2}(a). Thus, the drastic change in $V_\text{ISHE}/P_\text{abs}$ across 150 K at $f_0=4.0$ GHz can be attributed to the change in the magnetization dynamics in the Y$_3$Fe$_5$O$_{12}$ layer. In fact, by decreasing $T$, the microwave absorption intensity $P_\text{abs}$ decreased clearly across $T=150$ K as shown in Fig.~\ref{fig2}(c), suggesting the change of the magnetization dynamics in the Y$_3$Fe$_5$O$_{12}$ layer across $T=150$ K.

\begin{figure}[tb]
\includegraphics[scale=1]{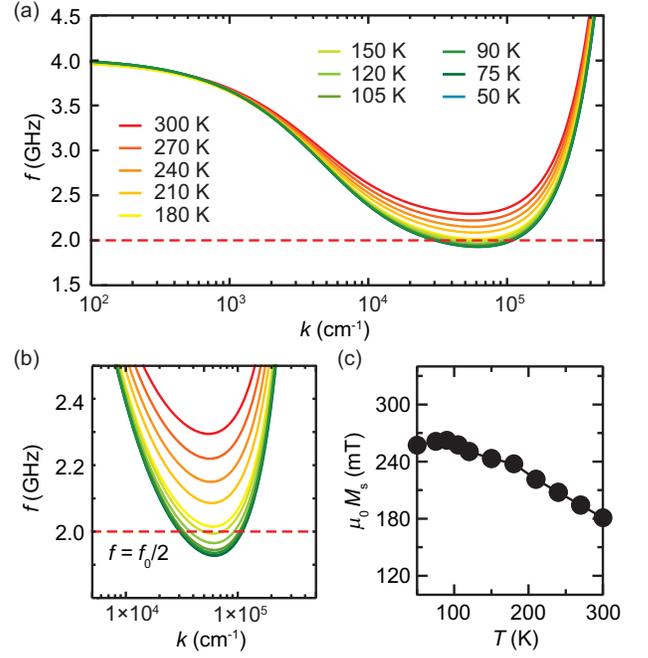}
\caption{{\bfseries Magnon dispersion.} (a) The lowest-energy branch of the magnon spectra for the Pt/Y$_3$Fe$_5$O$_{12}$ film calculated for the resonance condition at $f_0=4.0$ GHz. The dispersions were calculated using $\gamma=1.84\times 10^{11}$ Ts$^{-1}$. The dotted red line denotes $f=f_0/2=2.0$ GHz. (b) The magnified view of the lowest-energy branch of the magnon spectra.  (c) Temperature dependence of the saturation magnetization $M_\text{s}$ estimated from the resonance field data. 
}
\label{fig3} 
\end{figure}

\begin{figure*}[tb]
\includegraphics[scale=1]{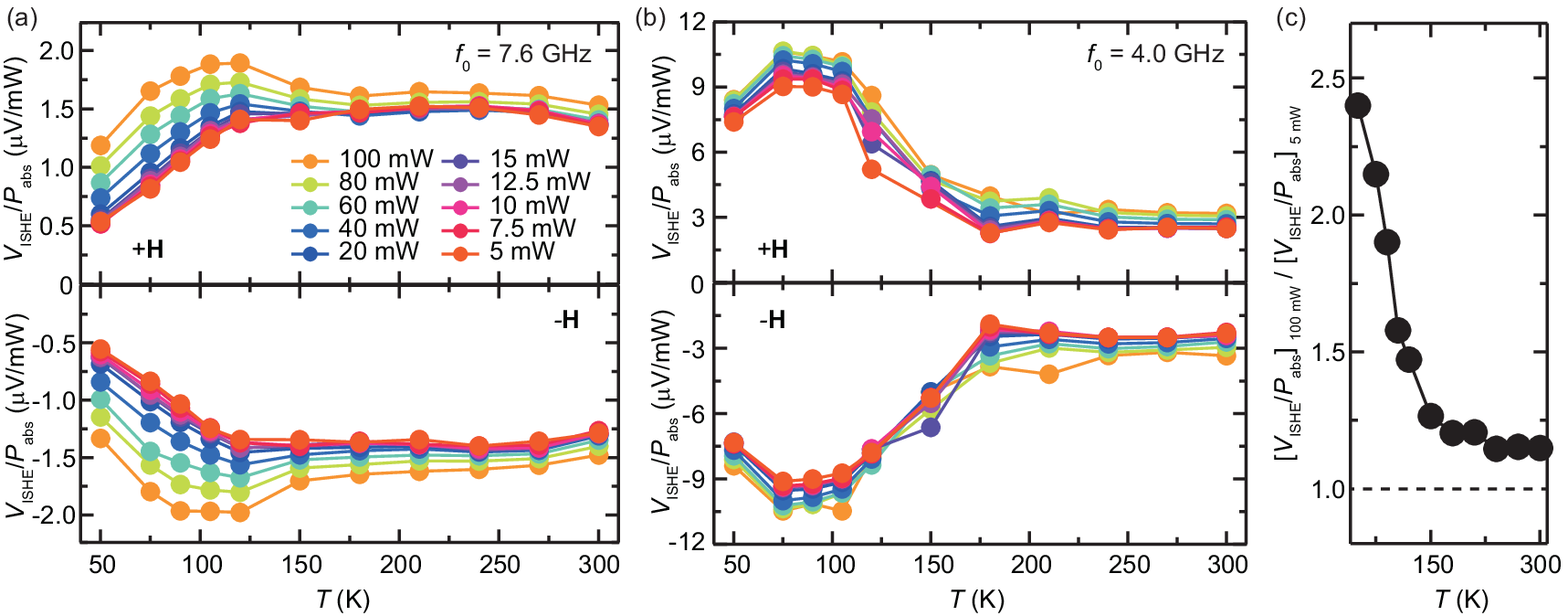}
\caption{{\bfseries Temperature evolution of spin-current emission for different microwave powers.} (a) Temperature $T$ dependence of $V_\text{ISHE}/P_\text{abs}$ at $f_0=7.6$ GHz for the in-plane magnetic field $\bf H$ (the upper panel) and reversed in-plane magnetic field $-\bf H$ (the lower panel). (b) $T$ dependence of $V_\text{ISHE}/P_\text{abs}$ at $f_0=4.0$ GHz for the in-plane magnetic field $\bf H$ (the upper panel) and $-\bf H$ (the lower panel). (c) $T$ dependence of $[V_\text{ISHE}/P_\text{abs}]_{100\text{ mW}}/[V_\text{ISHE}/P_\text{abs}]_{5\text{ mW}}$ at $f_0=7.6$ GHz. $[V_\text{ISHE}/P_\text{abs}]_{100\text{ mW}}$ and $[V_\text{ISHE}/P_\text{abs}]_{5\text{ mW}}$ are $V_\text{ISHE}/P_\text{abs}$ measured at $P_\text{in}=100$ mW and 5 mW, respectively.}
\label{fig4} 
\end{figure*}

The origin of the temperature-induced drastic change of the spin-conversion efficiency $V_\text{ISHE}/P_\text{abs}$ shown in Fig.~\ref{fig2}(a) is enhanced spin-current emission triggered by the three magnon splitting. The three-magnon splitting creates a pair of magnons with the opposite wavevectors and the frequency $f_0/2$ from the uniform magnon with $f_0$ [see also Fig.~\ref{fig1}(a)]. The splitting process redistributes the magnons and changes the relaxation rate of the spin system, increasing the steady-state angular momentum stored in the spin system, or resulting in the stabilized enhancement of the spin-current emission.~\cite{KurebayashiNM,SakimuraNC} The splitting is allowed only when $f_0/2>f_\text{min}$, where $f_\text{min}$ is the minimum frequency of the magnon dispersion, because of the energy and momentum conservation laws. This condition can readily be found by finding $f_\text{min}$ for the thin Y$_3$Fe$_5$O$_{12}$ film from the lowest branch of the dipole-exchange magnon dispersion for the unpinned surface spin condition:~\cite{Kalinikos}
\begin{equation}
f=\sqrt{\Omega \left(\Omega + \omega_M -\omega_M Q\right)}, \label{disp2}
\end{equation}
where $\Omega=\omega_H + \omega_M (D/\mu_0 M_\text{s}) k^2$, $\omega_H =\gamma \mu_0 H$, $\omega_M =\gamma \mu_0 M_\text{s}$, and $Q =  1-  \left[1-\exp{(-kL)}\right]/(kL)$. $D=5.2\times 10^{-13}$ Tcm$^{2}$ is the exchange interaction constant, $L=5$ $\mu$m is the thickness of the Y$_3$Fe$_5$O$_{12}$ layer, and $k$ is the wavenumber of the magnons. $\gamma=1.84\times 10^{11}$ Ts$^{-1}$ is the gyromagnetic ratio. In Figs.~\ref{fig3}(a) and \ref{fig3}(b), we show the lowest branch of the magnon dispersion at different temperatures for the Pt/Y$_3$Fe$_5$O$_{12}$ film, calculated using Eq.~(\ref{disp2}). For the calculation, we used the saturation magnetization $M_\text{s}$ at each temperature [see Fig.~\ref{fig3}(c)], estimated from the resonance field data with Kittel's formula. We assumed that $D$ is independent of temperature, as demonstrated in literature.~\cite{LeCrawWalker,0022-3719-20-8-013, SakimuraNC} Although $D$ can slightly depend on temperature,~\cite{heider1988note} the shape of the magnon dispersion is not sensitive to the small variation of $D$. Figures~\ref{fig3}(a) and \ref{fig3}(b) demonstrate that the minimum frequency $f_\text{min}$ decreases with decreasing temperature and the splitting condition $f_0/2>f_\text{min}$ is satisfied below $T=150$ K; the magnon redistribution is responsible for the enhancement of $V_\text{ISHE}/P_\text{abs}$. Thus, this result demonstrates that the enhanced spin-current emission can be induced not only by changing the excitation frequency or power of the magnetization dynamics, but also by changing temperature.

\begin{figure}[tb]
\includegraphics[scale=1]{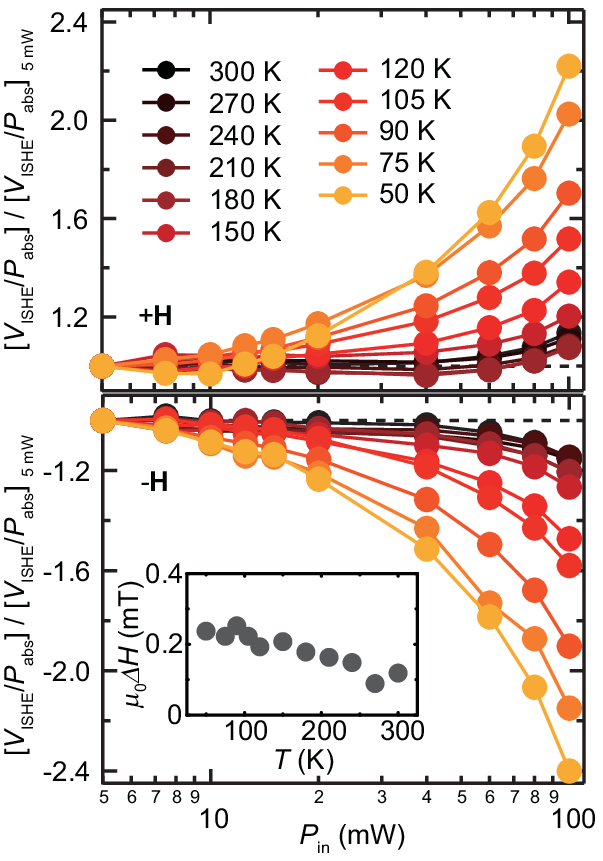}
\caption{{\bfseries Microwave power dependence of spin-current emission at different temperatures.} Microwave excitation power $P_\text{in}$ dependence of $[V_\text{ISHE}/P_\text{abs}]/[V_\text{ISHE}/P_\text{abs}]_{5\text{ mW}}$ at $f_0=7.6$ GHz for different temperatures. The in-plane magnetic field is $\bf H$ for the upper panel and $-\bf H$ for the lower panel, respectively. The inset shows $T$ dependence of the half-maximum full-width $\mu_0\Delta H$ of ferromagnetic resonance for the Pt/Y$_3$Fe$_5$O$_{12}$ film.}
\label{fig5} 
\end{figure}

Figures~\ref{fig4}(a) and \ref{fig4}(b) show temperature dependence of the spin-conversion efficiency $V_\text{ISHE}/P_\text{abs}$ at different microwave excitation powers $P_\text{in}$ for $f_0=7.6$ and 4.0 GHz, respectively. At $f_0=4.0$ GHz, the enhancement of $V_\text{ISHE}/P_\text{abs}$ due to the three-magnon splitting below 150 K is observed for all the excitation powers as shown in Fig.~\ref{fig4}(b). The drop in $V_\text{ISHE}/P_\text{abs}$ at $T=50$ K for $f_0=4.0$ GHz is induced by the decrease of the spin Hall conductivity shown in Fig.~\ref{fig2}(b); below 100K, the spin Hall conductivity, or the spin Hall angle, decreases with decreasing temperature, whereas the spin-current enhancement through the magnon splitting increases by decreasing temperature. The competition gives rise to the peak structure in $V_\text{ISHE}/P_\text{abs}$ around 70 K for 4.0 GHz. This result also shows that the enhancement factor is almost independent of the excitation power. In contrast, notably, the variation of $V_\text{ISHE}/P_\text{abs}$ depends on the excitation power, especially below 150 K, at $f_0=7.6$ GHz as shown in Fig.~\ref{fig4}(a). These features for $f_0=7.6$ and 4.0 GHz were confirmed in $V_\text{ISHE}/P_\text{abs}$ measured with the reversed external magnetic field [see the experimental data for $-{\bf H}$ in Figs.~\ref{fig4}(a) and \ref{fig4}(b)], indicating that the change of the spin-current emission from the magnetic insulator is responsible for the nontrivial behavior of $V_\text{ISHE}/P_\text{abs}$ at low temperatures.

To understand the temperature and power dependences of $V_\text{ISHE}/P_\text{abs}$ at $f_0=7.6$ GHz in details, we plot $[V_\text{ISHE}/P_\text{abs}]_{100\text{ mW}}/[V_\text{ISHE}/P_\text{abs}]_{5\text{ mW}}$ in Fig.~\ref{fig4}(c). For the spin-current emission in the linear magnetization dynamics regime, $V_\text{ISHE}/P_\text{abs}$ is constant with $P_\text{in}$, or $[V_\text{ISHE}/P_\text{abs}]_{100\text{ mW}}/[V_\text{ISHE}/P_\text{abs}]_{5\text{ mW}}=1$ because the emitted spin current is proportional to $P_\text{in}$.~\cite{AndoJAPfull} Since the three-magnon splitting is prohibited at $f_0=7.6$ GHz, $[V_\text{ISHE}/P_\text{abs}]_{100\text{ mW}}/[V_\text{ISHE}/P_\text{abs}]_{5\text{ mW}}\approx 1.2$, at $T=300$ K, demonstrates enhanced spin-current emission without the splitting of a pumped magnon.

The observed enhancement of the spin-current emission at $T=300$ K is induced by the four magnon scattering, where two magnons are created with the annihilation of two other magnons [see also Fig.~\ref{fig1}(a)].~\cite{PhysRevB.71.180411,PhysRevB.86.054414} The four-magnon scattering emerges at high microwave excitation powers $P_\text{in}>P_\text{th}$, known as the second order Suhl instability,~\cite{Suhl} where $P_\text{th}$ is the threshold power of the scattering. Although this process conserves the number of magnons, the magnon redistribution can decrease the relaxation rate of the spin system through the annihilation of the uniform magnons with large damping $\eta_0$ and creation of dipole-exchange magnons with small damping $\eta_{q}$. This results in the steady-state enhancement of the angular momentum stored in the spin system, or the enhanced spin-current emission.~\cite{SakimuraNC}  In the Pt/Y$_3$Fe$_5$O$_{12}$ film, the damping $\eta_0$ of the uniform magnon at low excitation powers is mainly dominated by the two-magnon scattering; the temperature dependence of the ferromagnetic resonance linewidth is almost independent of temperature as shown in the inset to Fig.~\ref{fig5}, indicating that the damping $\eta_0$  is not dominated by the temperature peak processes or the Kasuya-LeCraw mechanism.~\cite{ferro-relax} In contrast, the damping $\eta_q$ of the secondary magnons created by the four-magnon scattering is dominated by the Kasuya-LeCraw mechanism, since the two-magnon scattering events are suppressed due to the small group velocity; the group velocity of the secondary dipole-exchange magnons created at the same frequency as the uniform magnon can be close to zero because of the exchange-dominated standing spin-wave branches [see Fig.~\ref{fig1}(a)].~\cite{PhysRevB.70.224407,jungfleisch,PhysRevB.71.180411,chumak2014magnon} The exchange-dominated branches, i.e. the thickness modes, show the energy minimum not only at the bottom of the dispersion but also at the excitation frequency. Therefore, in the present system, the damping $\eta_0$ of the uniform magnon is dominated by the temperature-independent two-magnon scattering, whereas the damping $\eta_q$ of the secondary magnon is dominated by temperature-dependent three-particle confluences, such as the Kasuya-LeCraw process.~\cite{ferro-relax} In the presence of the four magnon scattering, the total number of the nonequilibrium magnons $N_t$ is expressed as~\cite{SakimuraNC}
\begin{equation}
\frac{N_t}{P_\text{abs}}=  \frac{1}{2\pi \eta_q \hbar f_0}\left[1-\frac{\chi''}{2\gamma M_\text{s}}(\eta_0-\eta_q)\right] \label{NtP},
\end{equation} 
where $\eta_q$ is defined as the average decay rate to the thermodynamic equilibrium of the degenerate secondary magnons for simplicity. The imaginary part of the susceptibility is expressed as 
\begin{equation}
\chi''=\frac{2\gamma M_\text{s}}{\eta_0+\eta_\text{sp}f(P_\text{in})},
\end{equation}
where 
\begin{equation}
f(P_\text{in})=\frac{1}{\sqrt{1-[\chi'' (\eta_0+\eta_\text{sp})/(2\gamma M_\text{s}) ]^4 (P_\text{in}/P_\text{th})^2}}. 
\end{equation}
Here, $\eta_\text{sp}$ is the decay constant of the uniform precession to degenerate magnons at $f_0$ due to scattering on sample inhomogeneities. Under the assumption that the spin-pumping efficiency is insensitive to the wavenumber $k$ of the nonequilibrium magnons, that is $V_\text{ISHE}\propto  j_s\propto N_t$, Eq.~(\ref{NtP}) is directly related to the spin-conversion efficiency: $V_\text{ISHE}/P_\text{abs}\propto N_t/P_\text{abs}$.

\begin{figure}[tb]
\includegraphics[scale=1]{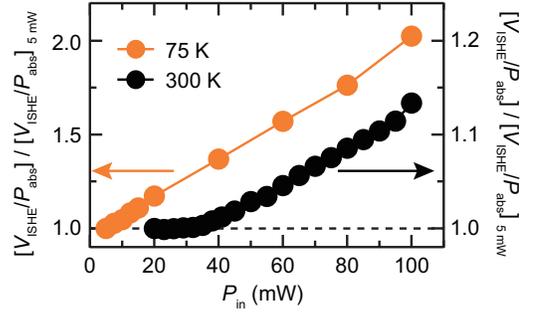}
\caption{{\bfseries Threshold power of spin-current enhancement.} Microwave excitation power $P_\text{in}$ dependence of $[V_\text{ISHE}/P_\text{abs}]/[V_\text{ISHE}/P_\text{abs}]_{5\text{ mW}}$ at $f_0=7.6$ GHz for $T=300$ K and $T=75$ K. }
\label{fig6} 
\end{figure}

The above model reveals that the spin-current enhancement due to the four-magnon scattering is responsible for the nontrivial behavior of the voltage generation shown in Fig.~\ref{fig4}(a). As shown in Fig.~\ref{fig4}(c), the nonlinearity of the spin-current emission is enhanced by decreasing temperature, from $[V_\text{ISHE}/P_\text{abs}]_{100\text{ mW}}/[V_\text{ISHE}/P_\text{abs}]_{5\text{ mW}}\approx 1.2$ at $T=300$ K to $[V_\text{ISHE}/P_\text{abs}]_{100\text{ mW}}/[V_\text{ISHE}/P_\text{abs}]_{5\text{ mW}}\approx 2.4$ at 50 K. Figure~\ref{fig5} shows microwave excitation power $P_\text{in}$ dependence of $V_\text{ISHE}/P_\text{abs}$ for $f_0=7.6$ GHz at different temperatures. This result clearly shows that the threshold power $P_\text{th}$ of the spin-current enhancement decreases with decreasing temperature, which is the origin of the nontrivial behavior of the temperature dependence of $V_\text{ISHE}/P_\text{abs}$ shown in Figs.~\ref{fig4}(a) and \ref{fig4}(c). The threshold power of the spin-current enhancement through the four-magnon process is very low at low temperatures, making it difficult to observe the threshold behavior. In fact, $V_\text{ISHE}/P_\text{abs}$ deviates from the prediction of the linear model even at the lowest microwave excitation power that is necessary to detect the ISHE voltage in the Pt/Y$_3$Fe$_5$O$_{12}$ film at $T=75$ K [see the orange circles in Fig.~\ref{fig6}]. At $T=300$ K, a clear threshold is observed around $P_\text{in}=40$ mW. The threshold power of the four-magnon scattering is given by~\cite{ferro-relax} $P_\text{th}\propto h_\text{th}^2=(\eta_0/\gamma)^2({2\eta_q/\sigma_q})$, where $h_\text{th}$ is the threshold microwave field and $\sigma_q$ is the coupling strength between the uniform and secondary magnons. For simplicity, we neglect the surface dipolar interactions, or $L\rightarrow \infty$. Under this approximation, the ferromagnetic resonance condition is given by $f_0= \gamma\mu_0 H$ and the coupling strength can be approximated as $\sigma_q=\gamma \mu_0 M_\text{s}$. Thus, the threshold power for the four-magnon scattering is proportional to 
\begin{equation}
h_\text{th}^2=\left(\frac{\eta_0}{\gamma}\right)^2 \left(\frac{2\eta_q}{\gamma \mu_0 M_\text{s}}\right). \label{threshold}
\end{equation}
Equation~(\ref{threshold}) predicts that the threshold power of the spin-current enhancement decreases with decreasing temperature, since $M_\text{s}$ increases by decreasing temperature as shown in Fig.~\ref{fig3}(c). Although the damping $\eta_0$ of the uniform magnon is almost independent of temperature as shown in the inset to Fig.~\ref{fig5}, the damping $\eta_q$ of the dipole-exchange magnon tends to decrease the threshold power, since $\eta_q$, dominated by the Kasuya-LeCraw process is approximately proportional to temperature.~\cite{ferro-relax} At high power excitations, the competition between the increase of the spin-current enhancement due to the four-magnon scattering and the decrease of the spin Hall effect by decreasing temperature gives rise to the peak structure in $V_\text{ISHE}/P_\text{abs}$ around 100 K for $f_0=7.6$ GHz [see Fig.~\ref{fig4}(a)].

In summary, we have demonstrated that the spin-current emission from a Y$_3$Fe$_5$O$_{12}$ film is strongly affected by nonlinear magnetization dynamics at low temperatures. The spin-current emission has been demonstrated to be enhanced even in the absence of the three-magnon splitting.~\cite{PhysRevB.86.134419} The experimental results presented in this paper are consistent with this result and further extend the physics of the nonlinear spin-current emission from the magnetic insulator. Our study reveals that the spin-current enhancement arises from both the three and four magnon scatterings depending on the excitation frequency and temperature. We show that the enhanced spin-current emission can be triggered by decreasing temperature, which is evidenced by our systematic measurements for the Pt/Y$_3$Fe$_5$O$_{12}$ film; the spin-current emission can be enhanced not only by changing the magnon excitation frequency or power, but also by changing temperature. This result demonstrates the generality of the crucial role of magnon interactions in the spin-current emission, combining the long-standing research on nonlinear spin physics with spintronics. 

This work was supported by JSPS KAKENHI Grant Numbers 26220604, 26103004, 26600078, PRESTO-JST ``Innovative nano-electronics through interdisciplinary collaboration among material, device and system layers," the Mitsubishi Foundation, the Asahi Glass Foundation, the Noguchi Institute, the Casio Science Promotion Foundation, and the Murata Science Foundation.

\end{document}